\documentclass[%
 reprint,
 amsmath,amssymb,
 aps,
]{revtex4-1}

\usepackage{graphicx}
\usepackage{dcolumn}
\usepackage{bm}
\usepackage{mathtools}
\usepackage{braket}
\usepackage{multirow}

\begin{document}

\preprint{APS/123-QED}

\title{High-speed and high-performance polarization-based quantum key distribution system without side channel effects caused by multiple lasers}

\author{Heasin Ko$^{\dagger}$, Byung-Seok Choi, Joong-Seon Choe, Kap-Joong Kim, Jong-Hoi Kim, and Chun Ju Youn}

\email{cjyoun@etri.re.kr, $^\dagger$seagod.ko@etri.re.kr}

\affiliation{Photonic/Wireless Convergence Components Research Division, Electronics and Telecommunications Research Institute, Daejeon, 34129, South Korea}
 
\date{\today}

\begin{abstract}
Side channel effects such as temporal disparity and intensity fluctuation of photon pulses caused by random bit generation with multiple laser diodes in high-speed polarization-based BB84 quantum key distribution (QKD) systems can be eliminated by increasing DC bias current condition. However, background photons caused by the spontaneous emission process under high DC bias current degrade the performance of the QKD systems. In this study, we investigated, for the first time, the effects of spontaneously emitted photons on the system performance in a high-speed QKD system at a clock rate of 400 MHz. Also, we further show improvements of system performance without side channel effects by utilizing temporal filtering technique with real-time FPGA signal processing.   
\begin{description}

\item[PACS numbers]
03.67.Dd
\end{description}
\end{abstract}

\pacs{03.67.Dd}
\maketitle


\section{Introduction}
 
Free-space quantum key distribution (QKD) system provides availability of unconditionally secure key exchanges between two distant parties without fiber network infrastructure. Polarization is normally adopted as a physical observable for free-space QKD system and it has been largely studied and demonstrated in diverse situations such as moving platform \cite{wang2013direct}, aircraft \cite{nauerth2013air}, long distance \cite{schmitt2007experimental}, and daylight condition \cite{liao2017long}. Recently, successful distribution of entangled photon pairs over a distance of 1200 km using quantum satellite \cite{Yin2017Satellite} and satellite to ground QKD \cite{Liao2017Satellite} were reported, which arouses expectations that unconditionally secure bit exchanges through a global network is feasible in the near future. However, such unconditional quantum security is only guaranteed under implementations where all components, both in the sender and receiver, are properly operated without any device loopholes \cite{ko2016informatic,lydersen2010hacking,nauerth2009information,nakata2017intensity,ko2017Critical}. 

\begin{figure*} [!t]
\centering
\includegraphics[width=\linewidth]{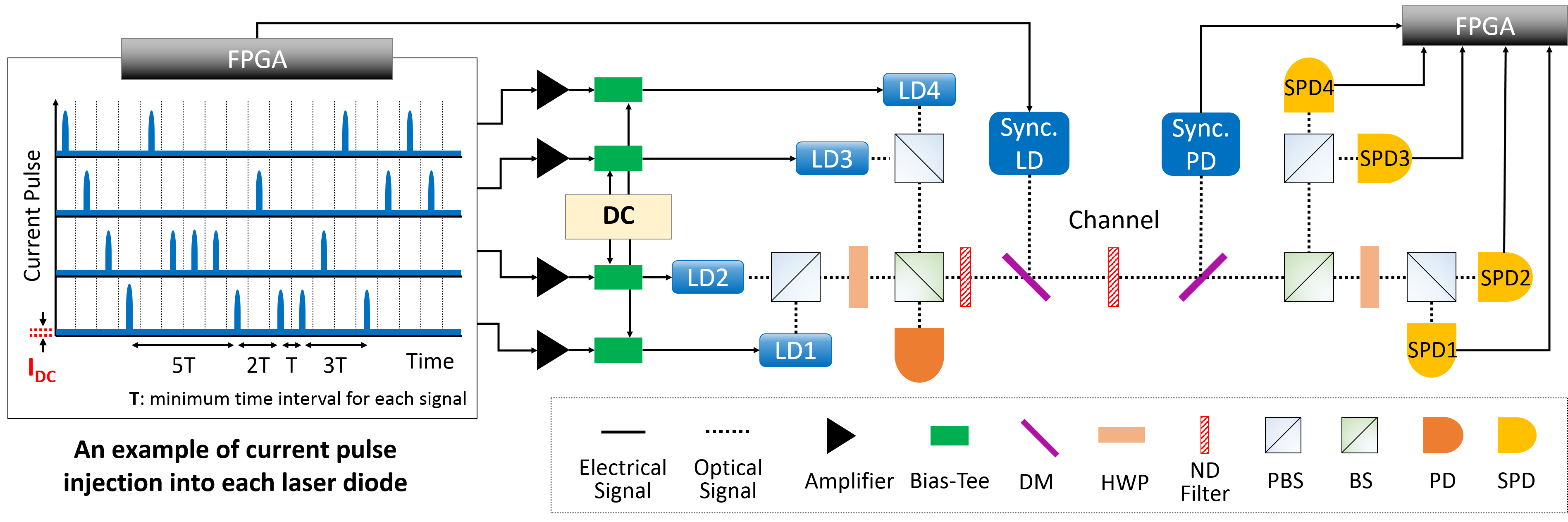}
\caption{\label{fig:setup} Experimental setup for measuring performance degradation of polarization-based BB84 QKD system under high DC bias current. An example of aperiodic current signals generated from FPGA is depicted. FPGA; field programmable gate array, LD; laser diode, Sync. LD; synchronization laser diode, Sync. PD; synchronization photodiode, DM; dichroic mirror, HWP; half wave plate, ND filter; neutral density filter, PBS; polarization beam splitter, BS; beam splitter, PD; photo-detector, SPD; single photon detector.}
\end{figure*}

In most free-space QKD systems, multiple semiconductor lasers with passive optics are utilized to randomly generate four different polarization states \cite{wang2013direct,nauerth2013air,schmitt2007experimental,liao2017long,Yin2017Satellite,Liao2017Satellite}. One of the polarization states can be transmitted by turning on one of the laser diodes exclusively for each time slot. We recently reported on side channel effects in random bit generation with multiple laser diodes in a polarization-based QKD system \cite{ko2017Critical}. In the above paper, we clearly showed that the temporal position and intensity of the photon pulses from each laser diode can be largely fluctuated according to the time interval between consecutive pulses from a single laser, the so called temporal disparity and intensity fluctuation. Although this issue is extremely critical in terms of the security of polarization-based QKD system with multiple laser diodes, it has not been clearly investigated because these effects are not severe for speeds of operation of 100MHz or lower, which are adopted in most representative free-space QKD demonstrations \cite{schmitt2007experimental,nauerth2013air,wang2013direct,liao2017long}. However, these side channel effects apparently occur in high-speed QKD systems, especially under the condition of low DC bias current injected into laser diodes, due to the dynamics of the initial carrier density and photon density. Since unconditional security is threatened by the side channel effects, such effects must be eliminated in the physical implementation, which can be accomplished by increasing the the DC bias current injected into the laser diodes. In this case, the performance of the QKD system can be unavoidably degraded due to the background photon noise from the spontaneous emission process of four laser diodes, especially for high DC bias current condition.

In this paper, we experimentally demonstrate, for the first time, the unavoidable performance degradation of high-speed polarization-based BB84 \cite{BB84} QKD system with multiple laser didoes to eliminate the side channel effects. First, we demonstrated that it is possible to achieve superior performance of the QKD system under zero DC bias condition even for a clock speed of 400 MHz, which is several times faster than previous representative free-space BB84 QKD demonstrations \cite{wang2013direct,nauerth2013air,schmitt2007experimental,liao2017long,Yin2017Satellite}. Second, we investigated how the side channels of temporal disparity and intensity fluctuation under zero DC bias current threaten the security of the system. Third, we repeated the QKD experiments under high DC bias current where the side channels are effectively closed. We quantitatively measured how the background photon noises caused by the spontaneous emission process under high DC bias current limits the performance of the quantum bit error rate (QBER) and secure key rate. Here, we showed that the photon counts by spontaneous emission of laser diodes can be a dominant factor that limits the performance of the high-speed system other than the dark counts of single-photon detectors. In addition, we demonstrated that the system performance can be significantly improved by using the temporal filtering technique. Finally, we discussed some strategies towards superior performance for high-speed QKD systems without the aforementioned side channel effects.

The remaining part of this paper is structured as follows. In sect. 2, the experimental setup to measure the effects of DC bias current on the performance of the QKD system is described. The experimental results of the QBER and key rates when side channel effects are neglected and eliminated are presented in sect. 3. In addition, improvements of the system performance by using the temporal filtering technique, as well as some analysis, are presented. In sect. 4, further discussions towards high-speed and high-performance polarization-based QKD system are presented. Some concluding remarks are presented in sect.5.


\section{Experimental Setup}

The experimental setup to measure the effects of different DC bias currents of laser diodes on the performance of QKD system is shown in Fig.~\ref{fig:setup}. We implemented a polarization-based BB84 QKD system with four semiconductor laser diodes and passive optics such as beam splitters, polarization beam splitters, half wave plates, and neutral density filters. We randomly injected electrical pulses to each laser diode as shown in Fig.~\ref{fig:setup}, which results in aperiodic current injection in terms of each laser diode. Single longitudinal mode vertical-cavity surface-emitting lasers (VCSELs) with a lasing wavelength of 787.5 nm were utilized for photon sources. We injected electrical current pulses of 200 ps full width at half maximum (FWHM) to each laser diode to generate optical pulses of 65 ps FWHM. Note that the optical pulse width is smaller than that of the electrical pulse injection due to the dynamics of large signal modulation with gain-switching method \cite{colren2012diode}. The amplitude of current pulses was controlled differently for different $I_{DC}$, such that we can generate similar optical pulses of 65 ps width where $I_{DC}$ is the DC bias current injected into the laser diodes as shown in the current pulse diagram in Fig.~\ref{fig:setup}. The photon pulses were generated at a clock rate of 400 MHz and attenuated to the mean photon number of 0.5. Note that clock speed can be increased to a higher speed than 400 MHz considering the fact that the width of the optical pulses is 65 ps. The clock signal for synchronization at the wavelength of 1550 nm was combined with the quantum signal using a dichroic mirror. A neutral density filter with 10 dB attenuation was added for emulating the channel loss.

The combined quantum signal and clock signal were split through another dichroic mirror in the receiver. The clock signal, which was recovered through a high-speed photo-detector, was processed at the FPGA to synchronize it with the quantum signal. Quantum signal detection was carried out with a four channel Si-APD based single photon detector (PerkinElmer SPCM-AQ4C) whose dark counts are lower than 500 counts/s per channel and detection efficiency is approximately 50\% at the wavelength of our laser diodes. The Bob's system loss, including optics loss and fiber coupling loss, was measured as 2 dB. The sifting process of raw keys and estimation of the QBER were conducted with FPGA-based real-time signal processing system. The QKD operation was carried out in a dark room condition to effectively eliminate other stray photons. Note that background noise photons caused by the spontaneous emission process of four laser diodes exist, which can differ according to the level of $I_{DC}$.   

\begin{figure}[t!]
\centering
\includegraphics[width=\linewidth]{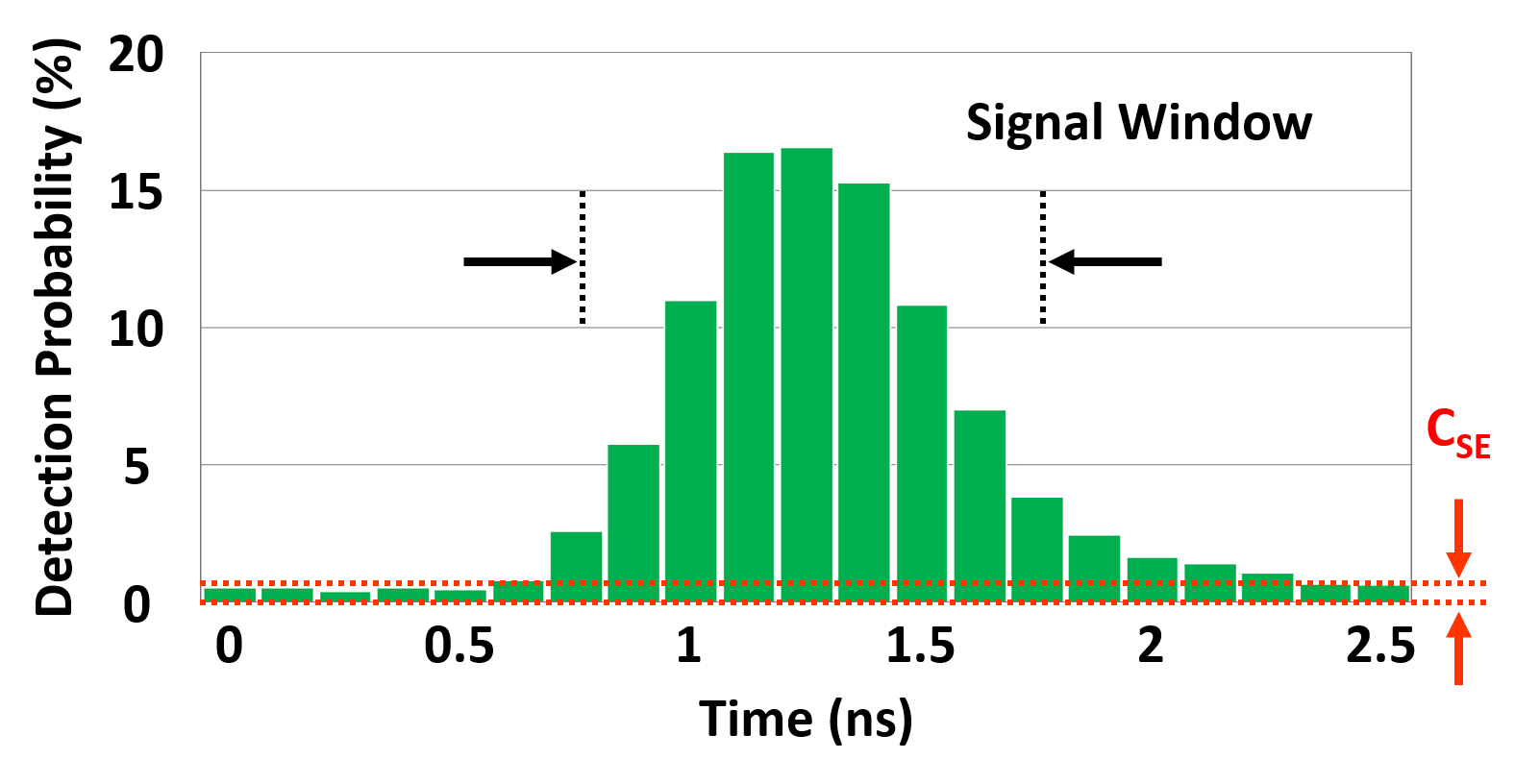}
\caption{\label{fig:prob_dist} Detection probability distribution of the received signals under $I_{DC}=0.95I_{th}$ which was estimated at the FPGA with a timing resolution of 125 ps. $\text{C}_{\text{SE}}$ represents the photon counts by spontaneous emission process.}
\end{figure}

Detection probability distribution of the received signals is depicted in Fig.~\ref{fig:prob_dist}. Even though the transmitted optical pulse width was approximately 65 ps, photon detection events were temporally broadened to approximately 750 ps FWHM. Note that detection probability distribution was unavoidably broader than the generated optical pulse width due to several factors, such as SPD response jitter, clock signal jitter, and differential digital data signal skews. In our system, most of the detection events were covered within the temporal region of 1.5 ns.

It is natural to set $I_{DC}$ as zero because any non-zero $I_{DC}$ generates spontaneously emitted photons for all temporal regions. In other words, we should set $I_{DC}$ as zero to turn on one of the four laser diodes exclusively for each time slot, otherwise, the four laser diodes always generate spontaneously emitted photons, which can definitely increase quantum bit error rates. Here, we measured QBER and key rates at $I_{DC}=0$, which is the condition in which the output photons from spontaneous emission are negligible. We repeated the experiment with $I_{DC}=0.95I_{th}$, which is the minimum $I_{DC}$ level of our VCSELs, to eliminate temporal disparity and intensity fluctuation at a clock rate of 400 MHz. For $I_{DC}=0.95I_{th}$, we investigated how the performance of the QKD system is degraded due to the detrimental effects of spontaneously emitted photons. We further attempted the temporal filtering technique limiting the signal time window of photon detection events to improve the system performance of the QBER and secure key rate. The valid signal window by temporal filtering was resized from 2.5 ns to 0.5 ns with the resolution of 0.25 ns to filter out background photon noises. 
 
\section{Experimental results}

\subsection{QKD operation without considering the side channel effects}

\begin{figure*}[t]
\centering
\includegraphics[width=\linewidth]{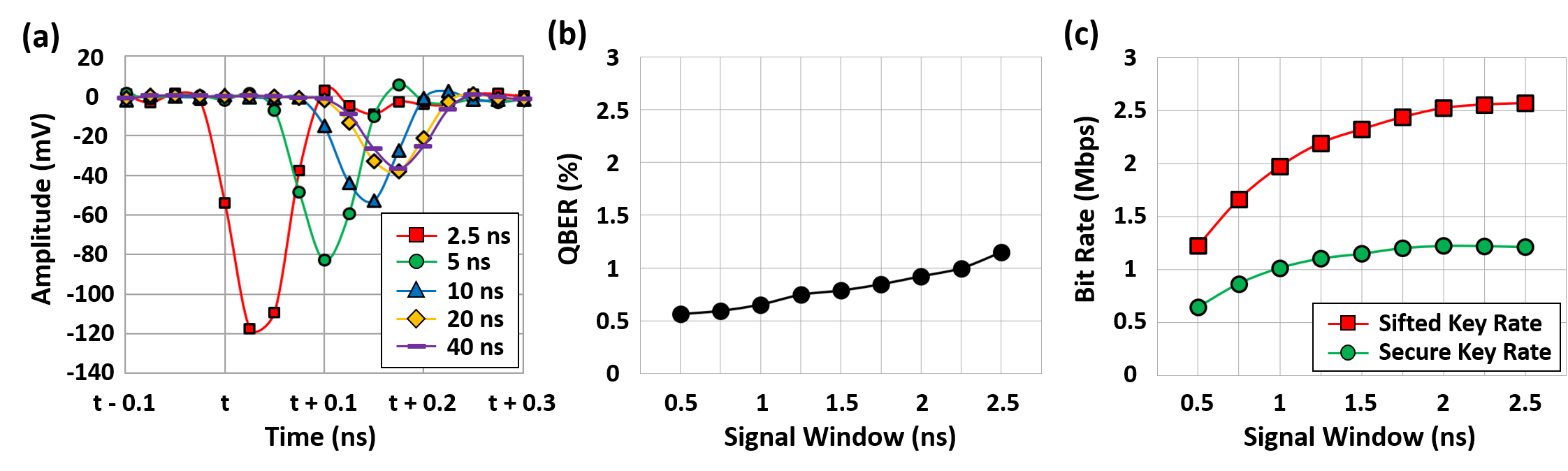}
\caption{\label{fig:performance1} Side channel effects and the performance of the QKD system under $I_{DC}=0$. (a) Side channel effects of temporal disparity and intensity fluctuation. Second pulses out of two consecutive pulses in a 1 \text{µ}s time block are measured for different time intervals from 2.5 ns to 40 ns. Details of the measurement methods are described in \cite{ko2017Critical}. (b) QBER as a function of signal window by temporal filtering. The signal window was resized from 2.5 ns to 0.5 ns with the resolution of 0.25 ns. (c) Sifted key rate and secure key rate as functions of the signal window by temporal filtering.}
\end{figure*}

The side channel effects which occurred in our system at a clock rate of 400 MHz for the condition with $I_{DC}=0$, are shown in Fig.~\ref{fig:performance1} (a). The optical pulses described in Fig.~\ref{fig:performance1} (a) represent the second pulses out of any two consecutive pulses for different time intervals from a single laser diode. The time position of the second pulses are depicted with respect to the current pulse injection to investigate temporal disparity among the second pulses. Here, time $t$ represents the expected time position of the output pulses. As described in \cite{ko2017Critical}, the pulses are temporally distinguishable depending on time intervals between two consecutive pulses, which can significantly endanger the system security if Eve utilizes previously known attack strategies accordingly \cite{Dusek2000Unambiguous,brassard2000limitations}. In addition, intensity fluctuation yields the degradation of security performance of BB84 QKD even for decoy-state QKD systems \cite{wang2008general,hayashi2014security,mizutani2015finite}. Thus, such side channels must be eliminated to guarantee the security of the QKD system. 

Any QKD system without considering such side channel effects may be wrongly interpreted as unconditionally secure one with superior QBER and key rates because the security loophole from such side channel effects are not reflected on the performance parameters such as QBER and sifted key rates. The performances of the QKD system for $I_{DC}=0$ are shown in Fig.~\ref{fig:performance1} (b) and Fig.~\ref{fig:performance1} (c). In this case, QBER of 1.04\% and key rate of 1.213 Mbps were achieved at a signal window of 2.5 ns even without implementing the temporal filtering technique. We verified that the QBER can be decreased to approximately 0.56\% by resizing the signal window down to 0.5 ns as shown in Fig.~\ref{fig:performance1} (b). Here, reducing signal window alleviates the effects of QBER caused by system jitter and clock drift. Also, detection events by dark counts of the SPD itself are eliminated as the signal window decreases. The impact of downsizing on the sifted key rates was imperceptible in the signal window range of 2 to 2.5 ns as shown in Fig.~\ref{fig:performance1} (c), because most signal photons exist within the range of 2 ns as shown in Fig.~\ref{fig:prob_dist}. Further downsizing yielded some reduction in the sifted keys, which became significant when the size was smaller than 1 ns. The secure key rate was calculated with the assumption of the decoy method \cite{hwang2003quantum,lo2005decoy} using the following simple equation \cite{ma2005practical}. 
\begin{equation}
R \approx q\bigg\{-\eta\mu f\big(e_{det}\big)H_2\big(e_{det}\big) + \eta\mu e^{-\mu}\Big[1-H_{2}\big(e_{det}\big)\Big]\bigg\}, \label{eq:secure_keyrate}
\end{equation}
where $q$ is the sifting ratio, $\mu$ is the mean photon number, $\eta$ is the transmittance, $e_{det}$ is the QBER, $f\big(e_{det}\big)$ is the error correction coefficient for given $e_{det}$, and $H_2\big(e_{det}\big)$ is the Shannon entropy where $H_2\big(e_{det}\big)=-e_{det}\text{log}_2(e_{det})-(1-e_{det})\text{log}_2(1-e_{det})$. Here, we used 1.22 for $f\big(e_{det}\big)$. The optimal window size in terms of the secure key rate was approximately 2.25 ns as shown in Fig.~\ref{fig:performance1} (c), where the secure key rate was recorded as 1.222 Mbps. Even though the system achieved superior performance parameters of QBER and key rates, however, the security of the QKD under $I_{DC}=0$ can be threatened due to the aforementioned side channel effects of temporal disparity and intensity fluctuation. 

\subsection{QKD operation considering the side channel effects}

\begin{figure*}[t!]
\centering
\includegraphics[width=\linewidth]{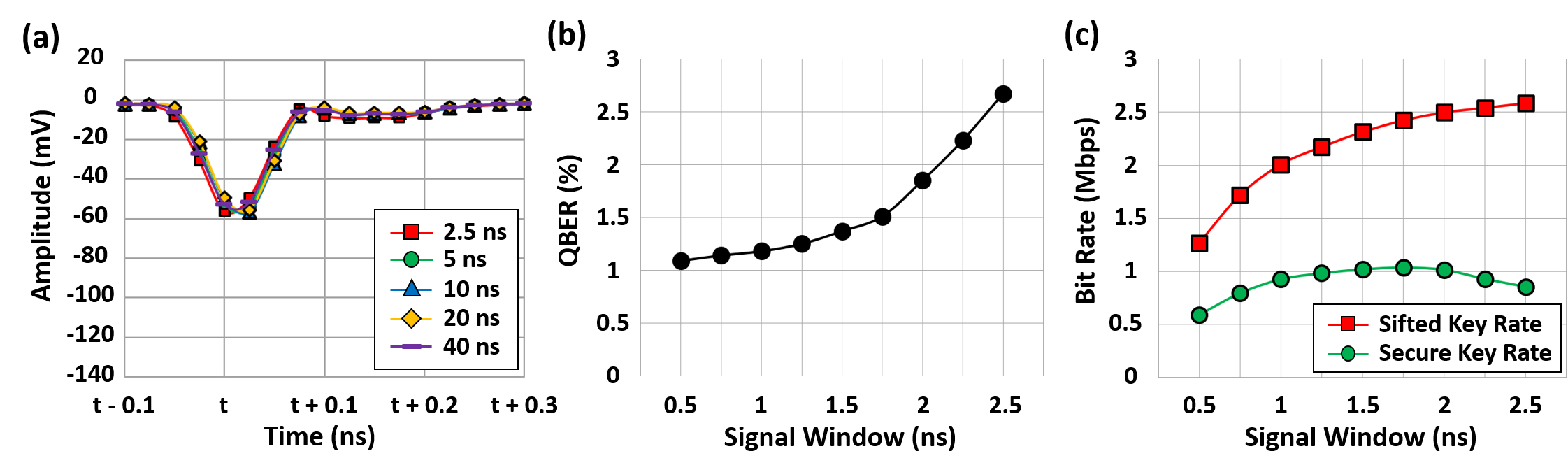}
\caption{\label{fig:performance2} Side channel effects and performance of the QKD system under $I_{DC}=0.95I_{th}$. (a) Side channel effects of temporal disparity and intensity fluctuation. Second pulses out of two consecutive pulses in a 1 \text{µ}s time block were measured for different time intervals from 2.5 ns to 40 ns. Details of the measurement methods are described in \cite{ko2017Critical}. (b) QBER as a function of signal window by temporal filtering. The signal window was resized from 2.5 ns to 0.5 ns with the resolution of 0.25 ns. (c) Sifted key rate and secure key rate as functions of the signal window by temporal filtering.}
\end{figure*}

The side channel effects were eliminated under $I_{DC}=0.95I_{th}$ as shown in Fig.~\ref{fig:performance2} (a). Pulses with different intervals between consecutive pulses are temporally overlapped and the intensity fluctuation becomes negligible. Thus, we can ensure that the side channels are eliminated under $I_{DC}=0.95I_{th}$. However, when $I_{DC}$ is increased close to the lasing threshold level $I_{th}$ of the laser diodes, the QBER can be increased owing to the enhanced amount of background noise photons due to the spontaneous emission process.
 
While the QBER was estimated around 1.04\% under $I_{DC}=0$, it was increased up to 2.67\% under $I_{DC}=0.95I_{th}$ as shown in Fig.~\ref{fig:performance2} (b). The excess QBER was mostly due to the spontaneously emitted noise photons from the four laser diodes, which were always generated even for unallowed time slots, due to the relatively high $I_{DC}$ as shown in the left part of Fig.~\ref{fig:setup}. Here, the spontaneously emitted noise photons under the $I_{DC}=0.95I_{th}$ condition become a dominant factor limiting the QBER performance, considering the fact that 1.63\% out of 2.67\% QBER is caused by the increase of $I_{DC}$ from $0$ to $0.95I_{th}$. 

The probability of the detection counts due to spontaneously emitted photons $\text{C}_{\text{SE}}$ are shown in Fig.~\ref{fig:prob_dist}. Here, temporal filtering evidently improved the QBER performance as the signal window decreases from 2.5 ns to 1.5 ns because $\text{C}_{\text{SE}}$ outside the signal window is effectively eliminated. In such signal windows, the sifted key rate was slightly diminished, which allowed a higher gain of the final secure key rate as shown in Fig.~\ref{fig:performance2} (c). For the signal windows smaller than 1.5 ns, however, the QBER performance was not significantly improved because temporal filtering reduces not only background photon noises, but also true signals, which significantly reduces the sifted key rate. Note that most of the received signals were covered within approximately 1.5 ns as shown in Fig.~\ref{fig:prob_dist}. For our QKD system, the measured optimal signal windows in terms of the final secure key rate for $I_{DC}=0.95I_{th}$ was approximately 1.75 ns, where the QBER was 1.51\% with the sacrifice of the detection events of 6.4\% . Here, the secure key rate was recorded as 1.037 Mbps. Further reduction of the signal window should be discouraged due to too much elimination of signal photons, which diminishes the final secure key rate. 


One can easily check that the condition of $I_{DC}=0.95I_{th}$ degrades the performance of QBER and key rate compared with $I_{DC}=0$ case. The final secure key rate without temporal filtering was decreased from 1.213 Mbps to 0.856 Mbps, which indicates a 29.4\% drop in the secure key rate. However, a high $I_{DC}$ condition is indispensable to close the side channels especially for a high-speed polarization based QKD system with multiple lasers. Under such condition, we can increase the secure key rate from 0.856 Mbps to 1.037 Mbps using temporal filtering, which indicates a 21.1\% improvement in the secure key rate. 


\section{Discussion}

\begin{figure}[t!]
\centering
\includegraphics[width=\linewidth]{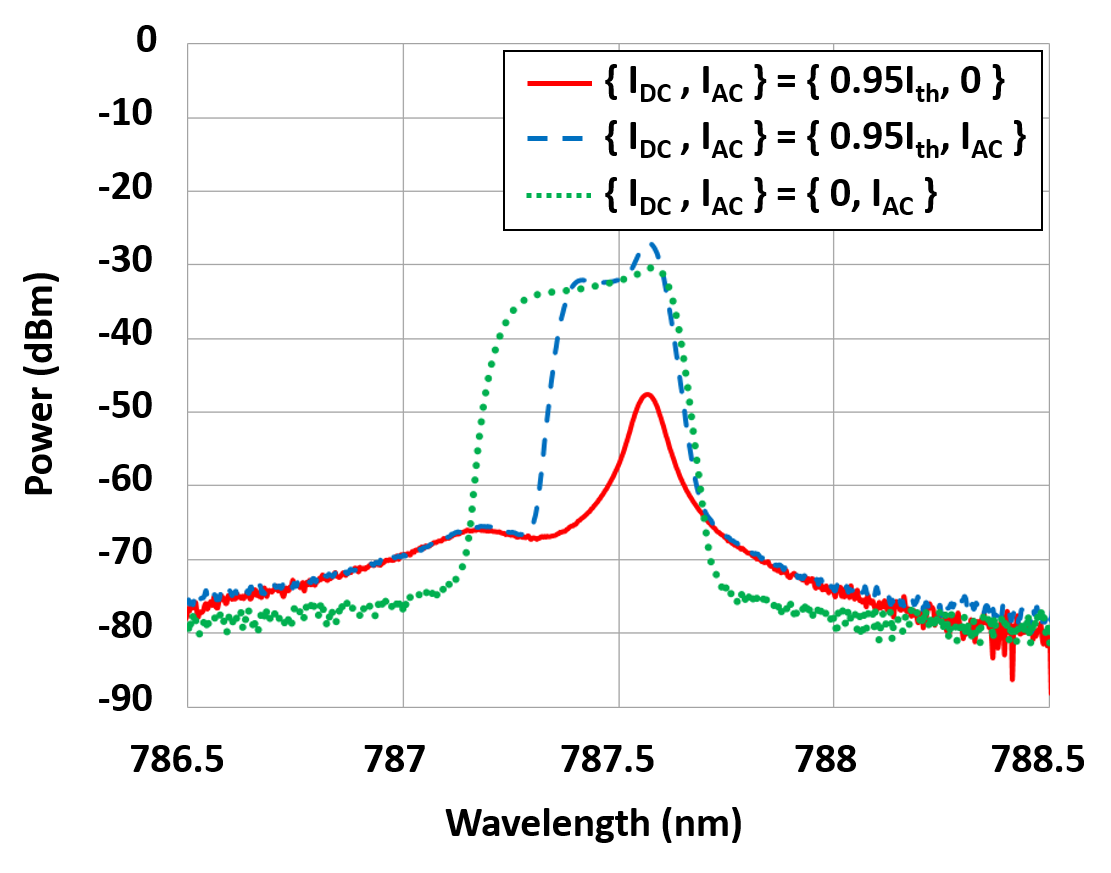}
\caption{\label{fig:spectrum} Spectral characteristics of photon pulses for different DC bias and AC pulse conditions.}
\end{figure}

Many previously known side channels of QKD systems can be easily avoided through simple countermeasures or simple monitoring methods without degradation of the system performance \cite{ko2016informatic,lydersen2010hacking,nauerth2009information,nakata2017intensity}. However, the temporal disparity and intensity fluctuation caused by random bit generation using multiple semiconductor laser diodes cannot be avoided without sacrificing the QBER performance, especially for high-speed BB84 QKD systems. Although the temporal filtering technique is definitely an effective method to improve the performance as demonstrated in the previous sections, there can be other possible strategies to alleviate the increase of the QBER as follows. 

One of the differences between spontaneously emitted and stimulated emitted photons is the spectral characteristics as shown in Fig.~\ref{fig:spectrum}. We measured the spectrum for three cases of \{$I_{DC},I_{AC}$\} for \{$0.95I_{th},0$\}, \{$0.95I_{th},I_{AC}$\}, and \{$0,I_{AC}$\}. Note that $I_{DC}=0.95I_{th}$ generates spontaneously emitted photons and AC pulse generates stimulated emitted photons. In a single longitudinal mode laser diode, most of the stimulated output photons are generated at a single wavelength as shown in Fig.~\ref{fig:spectrum}. On the other hand, photons by spontaneous emission shows broad spectral characteristics \cite{colren2012diode}. We can easily observe that the power of the sidebands is higher under $I_{DC}=0.95I_{th}$ than under $I_{DC}=0$. On the other hand, sidebands become negligible for $I_{DC}=0$ as shown in the dotted line in Fig.~\ref{fig:spectrum}. Therefore, we can eliminate sidebands caused by high $I_{DC}$ by utilizing an ultra-sharp spectral band pass filter, which mostly function as noise photons. 

Reducing system jitter is another strategy to improve the performance. The QBER caused by the spontaneous emission process can be continuously decreased as we carry out more temporal filtering, as shown in Fig.~\ref{fig:performance2} (a). If the system jitter including the SPD response jitter becomes smaller such that the received signal distribution becomes significantly narrower than that of our system, we can effectively decrease the QBER without sacrificing signal counts, which increases the final secure key rate. For our system, most of the jitter occurred at the stage of the single photon detection response. 
 
\section{Conclusion}
 
In this study, we experimentally demonstrated for the first time that the noise photons by spontaneous emission of multiple lasers, other than dark counts of the SPDs or system errors, can be a dominant factor limiting QBER performance in a high-speed of 400 MHz QKD operation considering the side channel effects of temporal disparity and intensity fluctuation. We showed that the performance of the secure key rate of the QKD system decreased from 1.213 Mbps to 856 kbps owing to the increase of QBER performance due to photon noise from the spontaneous emission process, which is non-negligible under 0.95$I_{th}$. We showed that the secure key rate of 856 kbps can be improved up to 1.037 Mbps by utilizing the temporal filtering technique, which should be considered as a desirable prerequisite for high-speed QKD systems with multiple laser diodes. In addition, we briefly discussed that using an ultra-sharp band pass filter and reducing system jitter can be effective techniques to achieve superior system performance. 
 
\section*{Funding Information}
This work was supported by Electronics and Telecommunications Research Institute (ETRI) grant funded by the Korean government. [Development of preliminary technologies for transceiver key components and system control in polarization based free space quantum key distribution]

\section*{acknowledgement}
The authors thank to Dr. Haeyoung Rha for helping data-processing with real-time FPGA system.



\nocite{*}

\begin{thebibliography}{99}

\bibitem{wang2013direct}
J. Y. Wang, B. Yang, S. K. Liao, L. Zhang, Q. Shen,  X. F. Hu, J. C. Wu, S. J. Yang, H. Jiang, Y. L. Tang, B. Zhong, H. Liang, W. Y. Liu, Y. H. Hu, Y. M. Huang, B. Qi, J. G. Ren, G. S. Pan, J. Yin, J. J. Jia,  Y. A. Chen, K. Chen, C. Z. Peng, and J. W. Pan, ``Direct and full-scale experimental verifications towards ground-satellite quantum key distribution,'' Nat. Photon. \textbf{7}, 387-393 (2013).

\bibitem{nauerth2013air}
S. Nauerth, F. Moll, M. Rau, C. Fuchs, J. Horwath, S. Frick, and H. Weinfurter, ``Air-to-ground quantum communication,'' Nat. Photon. \textbf{7}, 382-386 (2013).

\bibitem{schmitt2007experimental}
T. Schmitt-Manderbach, H. Weier, M. Fürst, R. Ursin, F. Tiefenbacher, T. Scheidl, J. Perdigues, Z. Sodnik, C. Kurtsiefer, J. G. Rarity, A. Zeilinger, and H. Weinfurter, ``Experimental demonstration of free-space decoy-state quantum key distribution over 144 km,'' Phys. Rev. Lett. \textbf{98}, 010504 (2007).

\bibitem{liao2017long}
S. K. Liao, H. L. Yong, C. Liu,	G. L. Shentu,	D. D. Li, J. Lin, H. Dai, S. Q. Zhao, B. Li, J. Y. Guan,	W. Chen,	Y. H. Gong, Y. Li, Z. H. Lin, G. S. Pan, J. S. Pelc,	M. M. Fejer, W. Z. Zhang, W. Y. Liu, J. Yin, J. G. Ren,	X. B. Wang, Q. Zhang, C. Z. Peng and J. W. Pan, ``Long-distance free-space quantum key distribution in daylight towards inter-satellite communication,'' Nat. Photon. \textbf{11}, 509–513 (2017).

\bibitem{Yin2017Satellite}
J. Yin, Y. Cao, Y. H. Li, S. K. Liao, L. Zhang, J. G. Ren, W. Q. Cai, W. Y. Liu, B. Li, H. Dai, G. B. Li, Q. M. Lu, Y. H. Gong, Y. Xu, S. L. Li, F. Z. Li, Y. Y. Yin, Z. Q. Jiang, M. Li, J. J. Jia, G. Ren, D. He, Y. L. Zhou, X. X. Zhang, N. Wang, X. Chang, Z. C. Zhu, N. L. Liu, Y. A. Chen, C. Y. Lu, R. Shu, C. Z. Peng, J. Y. Wang, and J. W. Pan, ``Satellite-based entanglement distribution over 1200 kilometers,'' Science \textbf{356}, 1140-1144 (2017).

\bibitem{Liao2017Satellite}
S. K. Liao, W. Q. Cai, W. Y. Liu, L. Zhang, Y. Li, J. G. Ren, J. Yin, Q. Shen,	Y. Cao, Z. P. Li,	F. Z. Li, X. W. Chen, L. H. Sun,	J. J. Jia,	J. C. Wu, X. J. Jiang, J. F. Wang,	Y. M. Huang, Q. Wang,	Y. L. Zhou, L. Deng, T. Xi, L. Ma, T. Hu, Q. Zhang, Y. A. Chen,	N. L. Liu, X. B. Wang, Z. C. Zhu,	C. Y. Lu, R. Shu,	C. Z. Peng, J. Y. Wang and J. W. Pan, ``Satellite-to-ground quantum key distribution,'' Nature \textbf{549}, 43-47 (2017).

\bibitem{ko2016informatic}
H. Ko, K. Lim, J. Oh, and J. K. K. Rhee, ``Informatic analysis for hidden pulse attack exploiting spectral characteristics of optics in plug-and-play quantum key distribution system,'' Quant. Inf. Proc. \textbf{15}, 4265-4282 (2016).

\bibitem{lydersen2010hacking}
L. Lydersen, C. Wiechers, C. Wittmann, D. Elser, J. Skaar, and V. Makarov, ``Hacking commercial quantum cryptography systems by tailored bright illumination," Nat. Photon. \textbf{4}, 686-689 (2010).

\bibitem{nauerth2009information}
S. Nauerth, M. Furst, T. Schmitt-Manderbach, H. Weier, and H. Weinfurter, ``Information leakage via side channels in freespace BB84 quantum cryptography,'' New J. Phys. \textbf{11}, 065001 (2009).

\bibitem{nakata2017intensity}
K. Nakata, A. Tomita, M.Fujiwara, K. I. Yoshino, A. Tajima, A. Okamoto, and K. Ogawa, ``Intensity fluctuation of a gain-switched semiconductor laser for quantum key distribution systems,'' Opt. Express \textbf{25}, 622--634 (2017).

\bibitem{ko2017Critical} 
H. Ko, B. S. Choi, J. S. Choe, K. J. Kim, J. H. Kim, and C. J. Youn, ``Critical side channel effects in random bit generation with multiple semiconductor lasers in a polarization-based quantum key distribution system,'' Opt. Express \textbf{25}, 20045--20055 (2017).

\bibitem{colren2012diode}
L. A. Coldren, S. W. Corzine, and M. L. Mashanovitch, Diode lasers and photonic integrated circuits (John Wiley and Sons, 2012).

\bibitem{BB84}
C. H. Bennett, and G. Brassard, ``Quantum cryptography: Public key distribution and coin tossing,'' in Proceedings of IEEE International Conference on Computers, Systems, and Signal Processing, 175--179 (1984).

\bibitem{hwang2003quantum}
W. Y. Hwang, ``Quantum key distribution with high loss: toward global secure communication,'' Phys. Rev. Lett. \textbf{91}, 057901 (2003).

\bibitem{lo2005decoy}
H. K. Lo, X. Ma, and K. Chen, ``Decoy state quantum key distribution,'' Phys. Rev. Lett. \textbf{94}, 230504.(2005).

\bibitem{ma2005practical}
X. Ma, B. Qi, Y. Zhao, and H. K. Lo, "Practical decoy state for quantum key distribution," Phys. Rev. A \textbf{72}, 012326 (2005).

\bibitem{Dusek2000Unambiguous}
M. Dusek, M. Jahma, and N. Lutkenhaus, ``Unambiguous state discrimination in quantum cryptography with weak coherent states,'' Phys. Rev. A \textbf{62}, 022306 (2000).

\bibitem{brassard2000limitations}
G. Brassard, N. Lutkenhaus, T. Mor, and B. C. Sanders, ``Limitations on practical quantum cryptography,'' Phys. Rev. Lett. \textbf{85}, 1330 (2000).

\bibitem{wang2008general}
X. B. Wang, C. Z. Peng, J. Zhang, L. Yang, and J. W. Pan, ``General theory of decoy-state quantum cryptography with source errors,'' Phys. Rev. A \textbf{77}, 042311 (2008).

\bibitem{hayashi2014security}
M. Hayashi, and R. Nakayama, ``Security analysis of the decoy method with the Bennett-Brassard 1984 protocol for finite key lengths," New J. Phys. \textbf{16}, 063009 (2014).

\bibitem{mizutani2015finite}
A. Mizutani, M. Curty, C. C. W. Lim, N. Imoto, and K. Tamaki, ``Finite-key security analysis of quantum key distribution with imperfect light sources,'' New J. Phys. \textbf{17}, 093011 (2015). 

\end{thebibliography}
\end{document}